\documentclass[fleqn,twoside]{article}
\usepackage{espcrc2}

\newcommand{\p}{\bot}

\newcommand{\dd}{\partial}
\newcommand{\de}{\delta}
\newcommand{\om}{\omega}
\newcommand{\Om}{\Omega}
\newcommand{\e}{\varepsilon}
\newcommand{\f}{\varphi}
\newcommand{\ls}{\left(}
\newcommand{\rs}{\right)}
\newcommand{\g}{\gamma}
\newcommand{\al}{\alpha}
\newcommand{\n}{\nu}
\newcommand{\La}{\Lambda}
\newcommand{\ra}{\rangle}
\newcommand{\ps}{\psi}
\newcommand{\te}{\theta}

\newcommand{\disn}[2]{$$\displaylines{\hfilneg\refstepcounter{equation}
            \label{#1} #2\hfilneg}$$}
\newcommand{\no}{\hfill\cr\hfill}
\newcommand{\nom}{\hfil \hskip 1em minus 1em (\theequation)}

\newcommand{\AmS}{{\protect\the\textfont2
  A\kern-.1667em\lower.5ex\hbox{M}\kern-.125emS}}

\title{On the correspondence between Light-Front Hamiltonian approach  and
Lorentz-covariant formulation for Quantum Gauge Theory}

\author{S.~A.~Paston\address{St.Petersburg State University,
        St.Petersburg, Russia},
        E.~V.~Prokhvatilov\addressmark
        and
        V.~A.~Franke\addressmark}

\begin{document}

\begin{abstract}
The problem of the restoring of the equivalence between Light-Front
(LF) Hamiltonian and conventional Lorentz-covariant formulations
of gauge theory is solved for QED(1+1) and (perturbatively to all
orders) for QCD(3+1). For QED(1+1) the LF Hamiltonian is constructed
which reproduces
the results of Lorentz-covariant theory. This is achieved
by bosonization of the model and by analysing the resulting bosonic
theory to all orders in the fermion mass.
 For QCD(3+1) we describe nonstandard
 regularization that allows to restore mentioned equivalence with
finite number of counterterms in LF Hamiltonian.

 \vspace{1pc}
\end{abstract}

\maketitle

\section{Introduction}
Light-Front (LF) Hamiltonian approach to Quantum Field Theory,
proposed by P.~Dirac \cite{dir}, uses, instead of usual Lorentz
coordinates $x^0,x^1,x^2,x^3$, the coordinate
$x^+=(x^0+x^3)/\sqrt{2}$ as a "time", and
 $x^-=(x^0-x^3)/\sqrt{2},\; x^{\p}=(x^1,x^2)$
as "spatial" coordinates. The quantization  surface is
$x^+=const$. Corresponding Hamiltonian $P_+=(P_0+P_3)/\sqrt{2}$
acts in LF Fock space. The "mathematical" vacuum in this space
coincides with physical vacuum state, defined by LF momentum
operator $P_-\ge0$. This simplification
 of the vacuum description essentially facilitates
nonperturbative eigenvalue problem for the $P_+$.

However there is general difficulty related with the singularity
at $p_-\to 0$, specific for LF formulation. So called naive
regularization of this singularity is achieved by the cutoff
$|p_-|\ge \e >0$. It cuts out zero modes ($p_-=0$) of fields and
breaks Lorentz and gauge symmetries. The other known
regularization, Discretized Light Cone Quantization (DLCQ), uses
the cutoff in the $x^-$ ($|x^-|\le L)$ plus periodic boundary
conditions for fields in $x^-$. Then the momentum $p_-$ becomes
discrete  ($p_-=p_n=\pi n/L$ with integer $n$) and zero modes
($p_-=0$) are expressed through nonzero modes via solving
canonical constraints. Gauge invariance can be kept in this
regularization scheme. However the solution of the constraints for
zero modes is very complicated due to the nonlinearity of these
constraints \cite{frnovpr}. Furthermore there is ordering problem
of quantum operators in these constraints.

Both mentioned regularizations break Lorentz invariance.
This can lead to a violation of the equivalence between LF  and
Lorentz-covariant formulations (after removing of the
regularization). The question about this
equivalence can be answered in perturbation theory. It was
supposed in \cite{bur} and proved to all orders in \cite{tmf97}
that for nongauge theories, like Yukawa model, it is sufficient
to modify only the
parameters, already present in naive LF Hamiltonian, to restore
this equivalence. It
was also found for these theories that the discussed equivalence
can be maintained without any modification of LF Hamiltonian if
one uses Pauli-Villars "ghost" fields for ultraviolet (UV)
regularization. A possibility to use LF nonperturbative
Hamiltonian approach with Pauli-Villars fields was also
investigated recently in \cite{brod}.

For gauge theories naively formulated on the LF (in the Light Cone
gauge, $A_-=0$) it is much more difficult to restore the
equivalence with Lorentz-covariant formalism. The comparison of
corresponding Feynman diagrams of LF and Lorentz-covariant
perturbation theories (for the simplest choices of UV
regularization) shows a difference between these perturbation
theories. This difference can not be compensated by adding finite
number of counterterms to the naive LF Hamiltonian \cite{tmf97}.
Only the complication of the regularization scheme by including
ghost fields (similar to Pauli-Villars ones) allows to restore the
equivalence with finite  (but large enough) number of such
counterterms in  LF Hamiltonian \cite{tmf99}.  We describe this
scheme in sect.~3.

The question about mentioned equivalence beyond the perturbation
theory in usual coupling constant can be answered for simpler
gauge field model like two dimensional QED (QED(1+1)), however,
only to all orders in different perturbation theory (in fermionic
mass). In this model we can bosonize the theory and analyze the
equivalent scalar field model within corresponding perturbation
theory (which is nonperturbative with respect to original coupling
constant).

Let us consider this more simple example of gauge theory at first,
and then return to more realistic gauge theories like QCD.

\section{QED(1+1) on the LF}
The QED(1+1), defined originally by the Lagrangian
 \disn{1}{
 L=-\frac{1}{4}F_{\mu\n}F^{\mu\n}+\bar\Psi(i\g^mD_\mu-M)\Psi,
 \nom}
can be transformed to its bosonized form \cite{hep}, described by
scalar field Lagrangian
 \disn{2}{
 L=\frac{1}{2}\ls\dd_\mu\Phi\dd^\mu\Phi-m^2\Phi^2\rs+\no
 +\frac{Mme^C}{2\pi}\cos(\te+\sqrt{4\pi}\Phi),
 \nom}
where $m=e/\sqrt{\pi}$ is Schwinger boson mass (the $e$ is
original coupling), $C=0.577\dots$ is Euler constant, and the $\te$ is
 "$\te$"-vacuum parameter. Here fermion mass $M$ plays the
role of the coupling in bosonized theory so that perturbation
theory in this coupling corresponds to chiral perturbation theory in
QED(1+1). The nonpolynomial form of scalar field interaction leads
in  perturbation theory to infinite sums of diagrams in each
finite order. It can be proved \cite{hep} that some partial sums
of these infinite sums are UV divergent in the 2nd order, whereas
for full (Lorentz-covariant) Green functions these divergencies
cancel (remaining only for vacuum diagrams).  Therefore physical
quantities are UV finite in this theory. Only at intermediate
steps of our analysis we need some UV regularization.

We compare LF and Lorentz-covariant perturbation theories for such
bosonized model using an effective resummation of
perturbation series in coordinate representation for Feynman
diagrams \cite{hep} and also using the methods of the paper
\cite{tmf97}. The results of this comparison can be formulated as
follows.

The difference between  considered perturbation theories can be
eliminated in the limit of removing regularizations if we use instead
of the  naive LF Hamiltonian
 \disn{3}{
 H\!=\!\!\!\int\!\! dx^-\!\!\ls\!\frac{1}{8\pi}m^2:\!\f^2\!:
 -\frac{\g}{2}e^{i\te}\!\!:\!e^{i\f}\!\!:-
 \frac{\g}{2}e^{-i\te}\!\!:\!e^{-i\f}\!\!:\!\rs\!\!,\no
 \g=\frac{Mme^C}{2\pi},\quad
 \f=\sqrt{4\pi}\Phi,\quad
 |p_-|\ge\e>0,
 \nom}
the "corrected" LF Hamiltonian:
 \disn{4}{
 H=\!\!\int\!\! dx^-\!\!\ls\frac{1}{8\pi}m^2:\!\f^2\!:
 -B:\!e^{i\f}\!:-B^*:\!e^{-i\f}\!:\!\rs\!-\no
 -2\pi e^{-2C}\frac{|B|^2}{m^2}
 \!\!\int\!\! dx^-dy^-\! \ls
 :e^{i\f(x^-)}e^{-i\f(y^-)}:-1\rs\!\!\times\no
 \times\te(|x^--y^-|-\al)
 \frac{v(\e(x^--y^-))}{|x^--y^-|}.
 \nom}
Here the terms, linear in $B$ and $B^*$  (new coupling constants),
are of the same form as in naive Hamiltonian, only  the term,
containing the $|B|^2$, is of new form (nonlocal in $x^-$). The
$\al$ is the UV regularization parameter, and the $v(z)$ is some
arbitrary continuous rapidly decreasing at the infinity function,
going to unity as $z\to 0$. The coupling $B$ can be perturbatively
written as a series in $\g$:
 \disn{5}{
 B=\frac{\g}{2}e^{i\te}+\sum_{n=2}^{\infty}\g^nB_n.
 \nom}
On the other side, it is related with the sum of all connected
"generalized tadpole" diagrams (i.e. diagrams with
external lines attached to only one vertex), which is described by
the "condensate" parameter
$A=\frac{\g}{2}\langle\Om|:e^{i(\f+\te)}:|\Om\ra$ of the
Lorentz-covariant formulation (the $|\Om\ra$ is physical vacuum
state in this formulation):
 \disn{5.1}{
 B+|B|^2w=A,
 \nom}
 \disn{6}{
 w=\frac{2\pi e^{-2C}}{m^2}\int dx^-\frac{\te(|x^-|-\e\al)}{|x^-|}v(x^-).
 \nom}
The eq.~(\ref{5.1}) can be solved with respect to the $B$:
 \disn{6.1}{
 B=-\frac{1}{2w}+\sqrt{\frac{1}{4w^2}+\frac{A'}{w}-A''^2}+iA'',
 \nom}
where $A=A'+iA''$, and the sign before the root respects the
perturbation theory. Within the perturbation theory in $\g$ one
cannot remove UV regularization ($\al\to 0$ and therefore
$w\to\infty$) in this expression due to UV divergencies of the
coefficients $B_n$. However, taking into account the validity of
the eq.~(\ref{5.1}) to all orders in $\g$, we can consider it
beyond the perturbation theory. Then we use the estimation for the
$A$ at $\al\to 0$ \cite{hep}:
 \disn{6.2}{
 A=\frac{\g^2}{4}w+const
 \nom}
and get for the $B$ in $\al\to 0$ limit UV finite result:
 \disn{7}{
 B=\frac{\g}{2} e^{i\hat\te(\te,M/e)}
 \nom}
so that all information about the condensate is contained in the phase
factor $e^{i\hat\te}$:
 \disn{8}{
 \sin\hat\te=2\frac{{\rm Im}A}{\g}=\langle\Om|:\sin(\f+\te):|\Om\ra.
 \nom}
 Then we can make a transformation, inverse to the
bosonization, but on the LF. Actually we need the expression only
for one independent component $\psi_+ $ of the bispinor field
$\ls\psi_+\atop\psi_-\rs$ due to the LF constraint, permitting to
write the $\psi_-$ in terms of $\psi_+$. One can use the exact
expression for the $\psi_+$ in terms of the $\f$ obtained in the
theory on the interval $|x^-|\le L$ with periodic boundary
conditions \cite{pirn,prok}. We need only to modify our corrected
bosonized theory by using discretized LF momentum $p_-$ instead of
continuous one and hence replacing the cutoff parameter $\e$ by
$\pi/L$. The necessary formulae for the $\psi_+$ has the following
form \cite{pirn,prok} (we choose here antiperiodic boundary
conditions for the fermion fields):
 \disn{9}{
 \psi_+(x)=\frac{1}{\sqrt{2L}}e^{-i \om}e^{-i\frac{\pi}{L}x^-  Q}
 e^{i\frac{\pi}{2L}x^-}:\!e^{-i\f(x)}\!:.
 \nom}
The operator $\om$ and the  charge operator $Q$ are canonically
conjugated so that the $\psi_+$ defined by the eq.~(\ref{9})  has
proper commutation relation with the charge. On the other side the
operator $e^{i\om}$  shifts Fourier modes $\psi_n$ of the field $\psi_+$
\cite{pirn,prok}:
 \disn{10}{
 e^{i\om}\psi_n e^{-i\om}=\psi_{n+1}.
 \nom}
If we separate the modes related with creation and annihilation
operators on the LF:
 \disn{11}{
 \psi_+(x)=\frac{1}{\sqrt{2L}}\Biggl(
 \sum_{n\ge 1}b_ne^{-i\frac{\pi}{L}(n-\frac{1}{2})x^-}+\no+
 \sum_{n\ge 0}d_n^+ e^{i\frac{\pi}{L}(n+\frac{1}{2})x^-}\Biggr),\quad
 b_n|0\ra=d_n|0\ra=0,
 \nom}
we can define the operator $e^{i\om}$  uniquely by specifying its
action on the LF vacuum $|0\ra$ as follows:
 \disn{12}{
 e^{i \om}|0\ra=b^+_1 |0\ra,\qquad
 e^{-i \om}|0\ra=d^+_0 |0\ra.
 \nom}
In such sense this operator is similar to a fermion.

We can now rewrite our corrected boson LF Hamiltonian in terms of
$\ps_+$ and  $e^{i\om}$. The result is remarkably simple:
 \disn{13}{
 H=\int\limits_{-L}^Ldx^-
 \biggl(\frac{e^2}{2}\ls \dd_-^{-1}
 [\psi_+^+ \psi_+]\rs^2-\frac{iM^2}{2}\times\no\times
 \psi_+^+\dd_-^{-1} \psi_+
 -\ls\frac{Me\:e^C}{4\pi^{3/2}} e^{-i\hat\te}\:e^{i \om}d_0^+
 +h.c.\rs\biggr).
 \nom}
This fermionic LF Hamiltonian differs from cano\-nical one (in
corresponding DLCQ scheme) only by last term, depending on zero
modes and vacuum condensate parameter $\hat\te$ which can be
related with chiral condensate by transforming the variables in
the eq.~(\ref{8}):
 \disn{14}{
 \sin\hat\te=-\frac{2\pi^{3/2}}{e\;e^C}
 \langle\Om|:\bar\Psi\g^5\Psi:|\Om\ra.
 \nom}
Our result can be formally reproduced if we modify (by proper
additional zero mode contribution) the constraint equation,
connecting the $\psi_-$ with   the $\psi_+$ on the LF. An analogous
modification of this constraint was got in the paper
 \cite{mac} where the method of
exact operator solution of massless Schwinger model was applied.

We  have some preliminary results of nonperturbative calculations
of the spectrum of lowest bound states with our  LF
Hamiltonian. These results agree with lattice calculations in
usual coordinates \cite{ham} for all values of coupling $M/e$
($\te =0$). This confirms the hope that our LF Hamiltonian is
valid nonperturbatively.

\section{QCD(3+1) on the LF}
Now let us discuss our results for (3+1)-dimen\-sional QCD
reported in \cite{tmf99}. They concern the question about the
equivalence of LF and Lorentz (and gauge) covariant  perturbation
theories for QCD. The main difficulty in the analysis of this
problem for Green functions results from additional pole
singularity at $p_-\to 0$ in the gluon propagator in Light Cone
gauge $A_-=0$ \cite{tmf97} (we use the Mandelstam-Leibbrandt
form):
 \disn{15}{
 \frac{-i\de^{ab}}{p^2+i0}\ls g_{\mu\n}-\frac{p_\mu n_\n+p_\n
 n_\mu}{2p_+p_-+i0}2p_+\rs.
 \nom}
The distortion of this pole due to LF cutoff $|p_-|\ge\e >0$ does
not disappear in the limit $\e\to 0$, and infinite number of new
counterterms are required to compensate this distortion
\cite{tmf97}. The simplest way to avoid this difficulty is to add
small mass-like parameter $\mu^2$ in the denominator:
 \disn{16}{
 \frac{1}{2p_+p_-+i0}\longrightarrow  \frac{1}{2p_+p_--\mu^2+i0},
 \nom}
and take the limit $\e\to 0$ before $\mu\to 0$). To describe this
modification with local Lagrangian we need to introduce ghost
fields $A'_{\mu}$ in addition to conventional $A_{\mu}$. We write
the free part of pure gluon Lagrangian as follows (using higher
derivatives and the parameter $\La$ for UV regularization):
 \disn{17}{
 L_0=-\frac{1}{4}\Biggl( f^{a,\mu\n} \ls 1+\frac{\dd^2}{\La^2}\rs
 f_{\mu\n}^a-\no
 -f'^{a,\mu\n}
 \ls 1+\frac{\dd^2}{\La^2}+\frac{2\dd_+\dd_-}{\mu^2}\Biggr)
 f'^a_{\mu\n}\rs,
 \nom}
where $f^a_{\mu\n}=\dd_\mu A^a_{\n}-\dd_\n A^a_{\mu}$,
$f'^a_{\mu\n}=\dd_\mu A'^a_{\n}-\dd_\n A'^a_{\mu}$ and
$A^a_-=A'^a_-=0$. Interaction terms depend only on summary field
$\bar A^a_{\mu}=A^a_{\mu}+A'^a_{\mu}$.

At fixed $\mu$ and $\La$  we get a theory with broken gauge
invariance but with preserved global  $SU(3)$-invariance. We put
into the Lagrangian all necessary interaction terms (but with
unknown
 coefficients) including those that are needed
for UV renormalization (these terms are local and can be taken in
Lorentz covariant form due to the restoring of this symmetry in
the $\e\to 0 ,\mu \to 0$ limit)
 \begin{eqnarray}
 L=L_0
 &\hskip -0.8em+\hskip -0.8em&
 c_0\dd_\mu\bar A^a_\n\dd^\mu\bar A^{a,\n}+
 c_1\dd_\mu\bar A^{a,\mu}\,\dd_\n\bar A^{a,\n}+\nonumber\\
 &\hskip -0.8em+\hskip -0.8em&
 c_2\bar A^a_\mu\bar A^{a,\mu}+
 c_3 f^{abc}\bar A^a_\mu\bar A^b_\n \dd^\mu\bar A^{c,\n}+\nonumber\\
 &\hskip -0.8em+\hskip -0.8em&
 \bar A^a_\mu\bar A^b_\n\bar A^c_\g\bar A^d_\de
 \bigl( c_4 f^{abe}f^{cde} g^{\mu\g}g^{\n \de}+\nonumber\\
 &\hskip -0.8em+\hskip -0.8em&
 \de^{ab}\de^{cd}\ls c_5 g^{\mu\g}g^{\n \de}+
 c_6 g^{\mu\n}g^{\g\de}\bigr)\rs.
 \label{18}
 \end{eqnarray}
In such a theory one can apply the methods of the paper
\cite{tmf97} to compare (at $\e\to 0$) the LF perturbation theory
and that taken in Lorentz coordinates within the same
regularization scheme.  We find that the difference between
mentioned perturbation theories can be compensated by changing of
the value of coefficient $c_2$ before the term of gluon mass form
$\bar A^a_\mu\bar A^{\mu,a}$ in naive LF Hamiltonian of this
theory. After that we can analyse further our regularized theory
in Lorentz coordinates and even make Euclidean continuation.

It is possible to prove by induction to all orders \cite{tmf99}
that in the limit $\mu\to 0$, $\Lambda\to\infty$ our theory can be
made finite and coinciding with the usual renormalized
(dimensionally regularized) theory in Light Cone gauge \cite{bas}
(for all Green functions).  To get this result one needs to choose
the unknown coefficients $c_i$ before all counterterms so that the
Green functions in each order coincided (after removing the
regularization) with those obtained in conventional dimensionally
regularized formulation and therefore satisfied Ward identities.
Besides, we need to correlate the limits $\mu\to 0$ and
$\La\to\infty$ to avoid infrared divergencies at $\mu\to 0$. It is
sufficient to take $\mu=\mu(\La)$ and to require that $\mu\La\to
0$ and $(\log\mu)/\La\to 0$.

Our resulting LF Hamiltonian for pure $SU(3)$ gluon fields
contains 7 unknown coefficients, including coefficient before
gluon mass term that takes also into account the difference
between LF and Lorentz coordinate formulations of our regularized
theory. The generalization of our scheme for full QCD with
fermions is described in \cite{tmf99}. In this case
there are 10 unknown coefficients in the LF Hamiltonian. We
hope that it is possible to find an analog of Ward identities
relating the coefficients $c_i$ at fixed $\La$. This problem seems
very important for our approach.

 \vskip 1em
{\bf Acknowledgements.} This work was supported by the Ministry of
Education of Russia (grant N~E00-3.3-316).

\end{document}